\documentclass[11pt,nofootinbib]{revtex4}
\usepackage{amsfonts,amssymb,amsmath,graphicx}
\usepackage{bm}
\usepackage[spanish,english]{babel}
\usepackage{graphicx}
\usepackage[latin1]{inputenc}
\usepackage{hyperref}

\begin{document}

\newcommand{\be}{\begin{equation}}
\newcommand{\ee}{\end{equation}}
\newcommand{\bea}{\begin{eqnarray}}
\newcommand{\eea}{\end{eqnarray}}
\newcommand{\nn}{\nonumber \\}
\newcommand{\e}{\mathrm{e}}

\title{Unification of the inflation with late-time acceleration \\ in Born-Infeld-$f(R)$ gravity}

\author{Andrey N. Makarenko  \\ {\small
Tomsk State Pedagogical University, ul. Kievskaya, 60, 634061 Tomsk,
Russia} \\ {\small
National Research Tomsk State University, Lenin Avenue, 36, 634050 Tomsk,
Russia}}

\begin{abstract}

We study accelerating dynamics from Born-Infeld-$f(R)$ gravity
in a simplified conformal approach without matter.
In  \cite{Eb7}  it was derived
 eventually any Dark Energy cosmology from above theory.
In this Letter we apply the technique of \cite{Eb7} to show that  Born-Infeld-$f(R)$
gravity  may describe very realistic universe admitting the unification of
early-time inflation with late-time acceleration.
Specifically, the evolution with periodic as well as non-periodic behavior is considered with possibility to cross the phantom-divide at early or late-times.

\end{abstract}


\maketitle
Various cosmological observations support
the current cosmic accelerated expansion.
To explain this phenomenon,
it is necessary to assume the existence of dark energy, which
has negative pressure, or propose that gravity is modified.
Its a very natural approach for the universe evolution
description where the early-time as well as late-time universe acceleration is
achieved by the modification of standard General Relativity \cite{review}. 
This leads to the fact that
unified description of the early-time inflation with late-time dark energy is natural in modified gravity as was
> first demonstrated by Nojiri-Odintsov \cite{R}
(for further unified models of
 this sort
 see refs \cite{Nojiri:2007as}). However,
the transition from decelerating phase to dark energy universe is
not yet well understood (possibly because it is not clear what is dark energy itself).

Recently, there was an interest for theories of gravity formulated in the Palatini formalism,  so-called Born-Infeld theory \cite{BI}.

The Palatini formulation brings
a number of restrictions and additional constraints to the metrics under
consideration so that it turns out to be quite difficult to get consistent
generalizations of the original  Born-Infeld  model. 

However, we have recently demonstrated that a non-perturbative and consistent generalization of Born-Infeld gravity is possible in the form of a Born-Infeld-$f(R)$ theory, which was introduced in ref.\cite{Eb6}. 

It was shown in ref. \cite{Eb7} that  Born-Infeld-$F(R)$ gravity without matter maybe easily reconstructed in conformal approach. In this way, it was derived
 eventually any Dark Energy cosmology from above theory.
In this Letter we apply the technique of \cite{Eb7} to show that  Born-Infeld-$f(R)$
gravity  may describe very realistic universe admitting the unification of
early-time inflation with late-time acceleration.
Specifically, the evolution with periodic as well as non-periodic behavior is considered with possibility to cross the phantom-divide at early or late-times.


In the
present letter we consider the Born-Infeld-$f(R)$ theory in conformal ansatz (see ref.\cite{Eb6, Eb7} for full details), it may be used to discuss a number of interesting situations.
In fact, we explicitly construct unification of the inflation with late-time acceleration using metrics proposed in the \cite{OD1}.

We now propose a modified action Born-Infeld containing an arbitrary function $f(R)$, where
$R=g^{\mu\nu} R_{\mu\nu}(\Gamma)$ \cite{Eb6}:
\be
\label{act}
S=\frac{2}{\kappa}\int
d^4x\left[\sqrt{|\det{\left(g_{\mu\nu}+\kappa
R_{\mu\nu}(\Gamma)\right)}|}-\lambda\sqrt{|g|}\right]
+\int d^4x\sqrt{|g|}f(R).
\ee

As noted earlier, we will work in the absence of matter, as soon as this case is realized in the conformal approach. 
Under the conformal approach we understand the situation, when the  metric $g_{\mu\nu}$  and   auxiliary metric (metric on which to build the Christoffel symbols, which are independent in Pallatini formalism)  are connected by transformation having the form of a conformal transformation ($g_{\mu\nu}=\Omega \, u_{\mu\nu}$).
  Varying the action  (\ref{act}) with respect to the connection we obtain the
equation
\be
\label{eq1}
\nabla_\alpha\left[\sqrt{q}\left( q^{-1}\right)^{\mu\nu}+\sqrt{g} g^{\mu\nu}
f_R\right]=0 \ ,
\ee
where 
\be \label{q1}
q_{\mu\nu}=g_{\mu\nu}+\kappa
R_{\mu\nu}(\Gamma)
\ee
and
$f_R\equiv df/dR$. The corresponding equation obtained by variation over the metric has the form
\be
\label{e1_1}
\sqrt{q}\left(q^{-1}\right)^{\mu\nu}-\lambda
\sqrt{g}g^{\mu\nu}+\frac{\kappa}{2}\sqrt{g}g^{\mu\nu} f(R)-\kappa \sqrt{g}
f_R R^{\mu\nu}=0.
\ee

Since we work conformal approach, it is sufficient to require the fulfillment of the following conditions:
\be
\label{q1}
q_{\mu\nu}=k(t) g_{\mu\nu}.
\ee

In this case we have an auxiliary metric $u_{\mu\nu}$  which defines the
covariant derivative and hence the Christoffel symbols
\be
\label{metric2}
\Gamma^\alpha_{\mu\nu}=\frac{1}{2} u^{\alpha\beta}\left(\partial_\mu
u_{\nu\beta}+\partial_\nu u_{\mu\beta}-\partial_\beta u_{\mu\nu}\right).
\ee

Here
\be
\label{umn}
u_{\mu\nu}=(k(t)+f_R)g_{\mu\nu}.
\ee

For the condition (\ref{q1})  together with the definition $q_{\mu\nu}$ it is
clear that the scalar curvature must also be proportional to the metric
$g_{\mu\nu}$. One can write the relationship between the scalar curvature and the metric as
\be
\label{Ruq}
R_{\mu\nu}=\frac{1}{\kappa}(k(t)-1)g_{\mu\nu}.
\ee
Let us consider  the spatially-flat FRW universe with metric
\be
\label{FRW}
ds^{2}=-dt^{2}+a^{2}(t)(dx^{2}+dy^{2}+dz^{2})\  .
\ee

The auxiliary metric will be given by the expression (\ref{umn}).


Let us denote the function connecting main and auxiliary metrics as  $u(t)=k(t)+f_R$. Suppose now that
$R_{\mu\nu}=r(t) g_{\mu\nu}$ where $r(t)$ is easy to find from the
Eq.(\ref{Ruq}). 
Finally, the equations take the following form
\begin{eqnarray}
r(t)&=&3H^2+\frac{3H\dot{u}}{u(t)}+\frac{3\dot{u}(t)^2}{4u(t)^2},\\
2\dot{H}&=&H\frac{\dot{u}(t)}{u(t)}+\frac{3\dot{u}(t)^2}{2u(t)^2}-\frac{\ddot{u}(t)^2}{u(t)}.
\end{eqnarray}
Here $H$ is Hubble rate ($H=\frac{\dot{a}}{a}$).
From these equations it is easy to get
\be
\label{con1}
u(t)=c \,\,r(t)
\ee
where $c$ is a constant. The remaining equations lead us to
\be
\label{eeq1}
H=\pm\sqrt{\frac{u}{3c}}-\frac{\dot{u}}{2u}.
\ee
From this result, as shown in \cite{Eb6}, the form of the function $f(R)$ maybe found explicitly
\be
\label{ff1}
f(R)=\frac{2}{\kappa}(\lambda-1)-R+\frac{c-\kappa}{8}R^2.
\ee


\subsection{Periodical behavior}

One can consider several examples which is demonstrated how transitions between phantom and non-phantom phases occur and that two phases are smoothly connected with each other \cite{OD1}.
Hence, there occurs the universe which contains several phantom phases corresponding to early time inflation and late time acceleration. In between phantom phases will be the standard non-phantom cosmology (radiation/matter dominated, expanding or shrinking one).
In oscillating universe there may appear several phantom/non-phantom transitions.

Let us consider the universe with the scale factor of the following form
\be
\label{eqq}
a=a_0e^{{h_0} t-\frac{t{h_1} \text{Cos}\left( \nu\, t\right)}{\nu}},
\ee
where $a_0$, $h_0$, $h_1$ and $\nu$ - are constants.
Then the Hubble parameter takes the form
\be
H=
h_0+{h_1} \text{Sin}\left( \nu\, t\right)
\ee
Let us now will look like relationship between metrics $g_{\mu\nu}$ and $u_{\mu\nu}$, it is necessary to find solutions to the equation (\ref{eeq1}):
\be
\label{Osts1}
u(t)=\frac{4 e^{-2 {h_0} t+\frac{2 {h_1} \text{Cos}(\nu\, t)}{v}}}{\left(2 c\pm\int_1^t \sqrt{2} e^{\frac{{h_1} \text{Cos}(\nu\, x)}{v}-{h_0} x} \, dx\right)^2},
\ee
note that for further discussion we will choose the plus sign in the expression (\ref{eqq}), since it suffices to choose negative constant $c$ to obtain a second case.

The figure 1 illustrates  the behavior of the scale factor (green line), the Hubble rate  (blue line), function $u(t)$ (red line) and the effective parameter  EoS (black line), which is defined by the following expression 
\be
w_{eff}=-1-\frac{2\dot{H}}{3H^2}.
\ee

When $h_0> h_1>0$, $H$ is always positive and the universe is expanding. Since
\be
\dot{H}=h_1\nu \text{Cos}(\nu\,t),
\ee
when
$h_1 \nu >0$, $w_{eff}$ is greater than $-1$ (non-phantom phase) when 
$\left(2n-1/2\right)\pi<\nu\,t<\left(2n+1/2\right)\pi$, and less than $-1$ (phantom phase) when $\left(2n+1/2\right)\pi<\nu\,t<\left(2n+3/2\right)\pi$ \cite{OD1}.

Thus, in our model occur oscillations  between the phantom and non phantom phases. Furthermore, it is clear that the derivative of the Hubble rate vanishes when $\nu\,t=\pi(n+1/2)$ and we get a situation where the Hubble rate becomes constant ($\dot H=0$).  We have several stages with an effective de Sitter solution for different values of the scalar curvature. One of these stages can be interpreted as inflationary, and as the next phase of accelerated expansion.


\begin{figure}[!h]
\begin{minipage}[h]{0.4\linewidth}
\includegraphics[angle=0, width=1\textwidth]{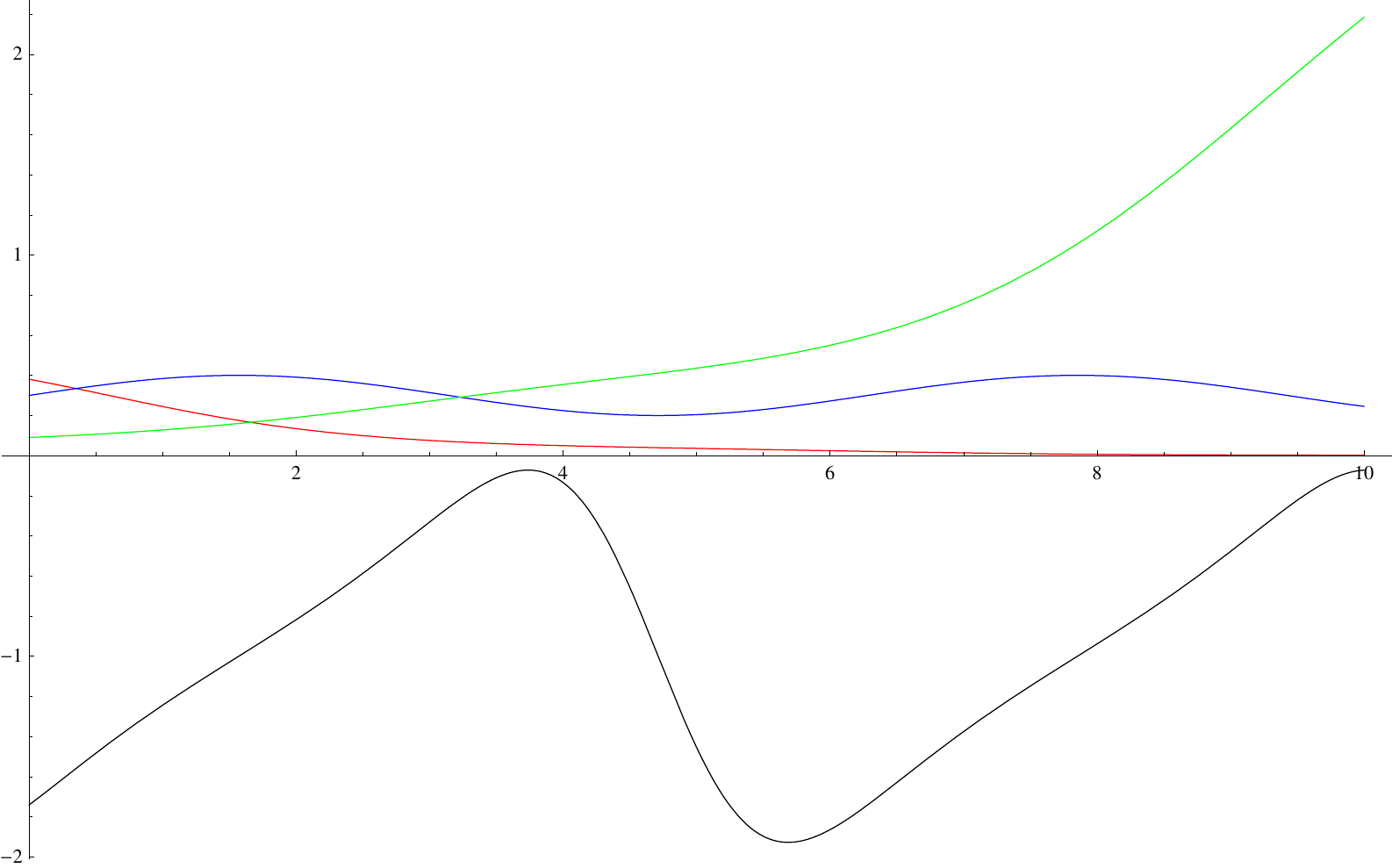}
\caption{scale factor (green line), the Hubble rate  (blue line), function $u$ (red line) and the effective parameter  EoS (black line) as a functions of $t$
with the parameters
$h_0=0.3$, $h_1=0.1$, $c=-5$, $\nu=1$ and $a_0=0.1$.}
\end{minipage}
\hspace{10mm}
\begin{minipage}[h]{0.4\linewidth}
\includegraphics[angle=0, width=1\textwidth]{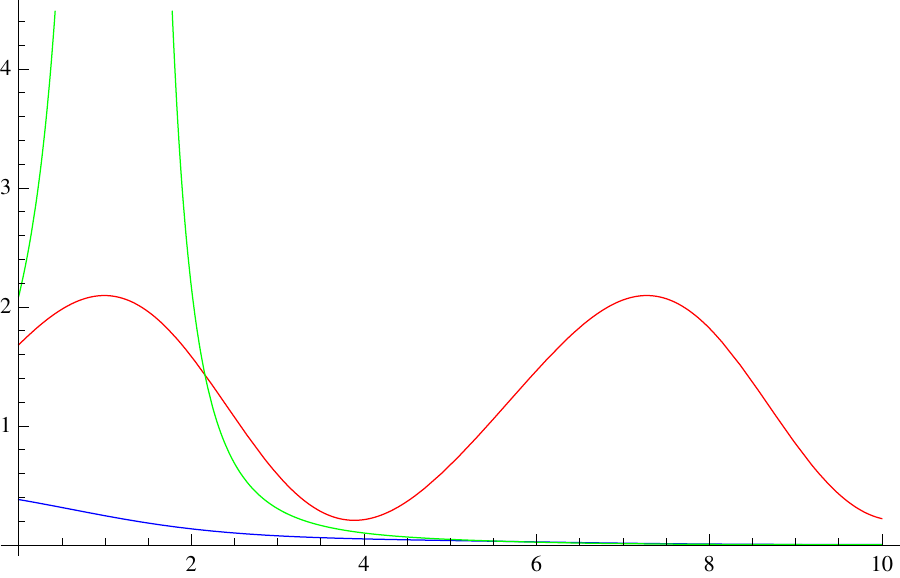}
\caption{ $u$  (green line) for  $h_0 =0.3$, $c= -0.1$, $h_1 = 0.1$ and $\nu=1$, 10$u$ (blue line) for  $h_0 =0.3$, $c= -5$, $h_1 = 0.1$ and $\nu=1$ and the scalar curvature constructed from the metric (\ref{eqq}) (red line) for $h_0 =0.3$, $h_1 = 0.1$ as a functions of $t$.}
\end{minipage}
\label{graph_1}
\end{figure}

In our model, the scalar curvature is determined by auxiliary metrics and it will be proportional to the function $u(t)$. Behavior of the scalar curvature is 
illustrated in Figure 2: the green line is a $u(t)$ for the following values of the constants -- $h_0 =0.3$, $c= -0.1$, $h_1 = 0.1$ and $\nu=1$; blue line is 
the graph of $10 u(t)$ for  $h_0 =0.3$, $c= -5$, $h_1 = 0.1$ and $\nu=1$; and red line is a graph of the scalar curvature constructed from the metric (\ref{eqq}). We see that in our model, the scalar curvature at large times tend to zero, and for a particular choice of constant have singularity. Whereas  the usual theory the scalar curvature constructed from the metric is a periodic function.

\subsection{Non Periodically behavior}
Consider another example illustrating the transition between the phantom and non phantom phases.

Suppose now that the metric has the form
\be
\label{eaa1}
a=a_0\left( \frac{t}{t_0-t}\right)^h_0,
\ee
where $a_0$, $h_0$ and $t_0$ are constants.
For this metric the Hubble rate have the next form
\be
H={h_0}\left(\frac{1}{t}+\frac{1}{{t_0}-t}\right).
\ee

From the equation (\ref{eeq1}) we obtain the next form of function $u$
$$u=\frac{4 ({h_0}-1)^2 \left(1-\frac{t}{t_0}\right)^{2 {h_0}} (t_0-{t})^{2 h_0}}{\left(2 ({h_0}-1) t^{{h_0}} 
\left(1-\frac{t}{t_0}\right)^{{h_0}} c
\pm t (t_0-{t})^{{h_0}} \text{Hypergeometric2F1}\left[1-{h_0},-{h_0},2-{h_0},\frac{t}{{t_0}}\right]
\right)^2},$$
where
$ \text{Hypergeometric2F1}\left[a,b,c,z\right]$ is the hypergeometric function $_2F_1(a,b;c;z)$. Again, we shall assume that in this expression selected plus sign.
It is easy to find
$$w_{eff}=-\frac{4 t-2 {t_0}+3{h_0} {t_0}}{3{h_0} {t_0}}$$
and draw a graph fig.3  illustrating this model.
When $0<t<t_0/2$, the universe is in non-phantom phase but when $t_0/2< t < t_0$, it is in phantom phase.  Hence, again the
unified phantom inflation/acceleration universe may emerge.

\begin{figure}[tbp]
\includegraphics[angle=0, width=0.5\textwidth]{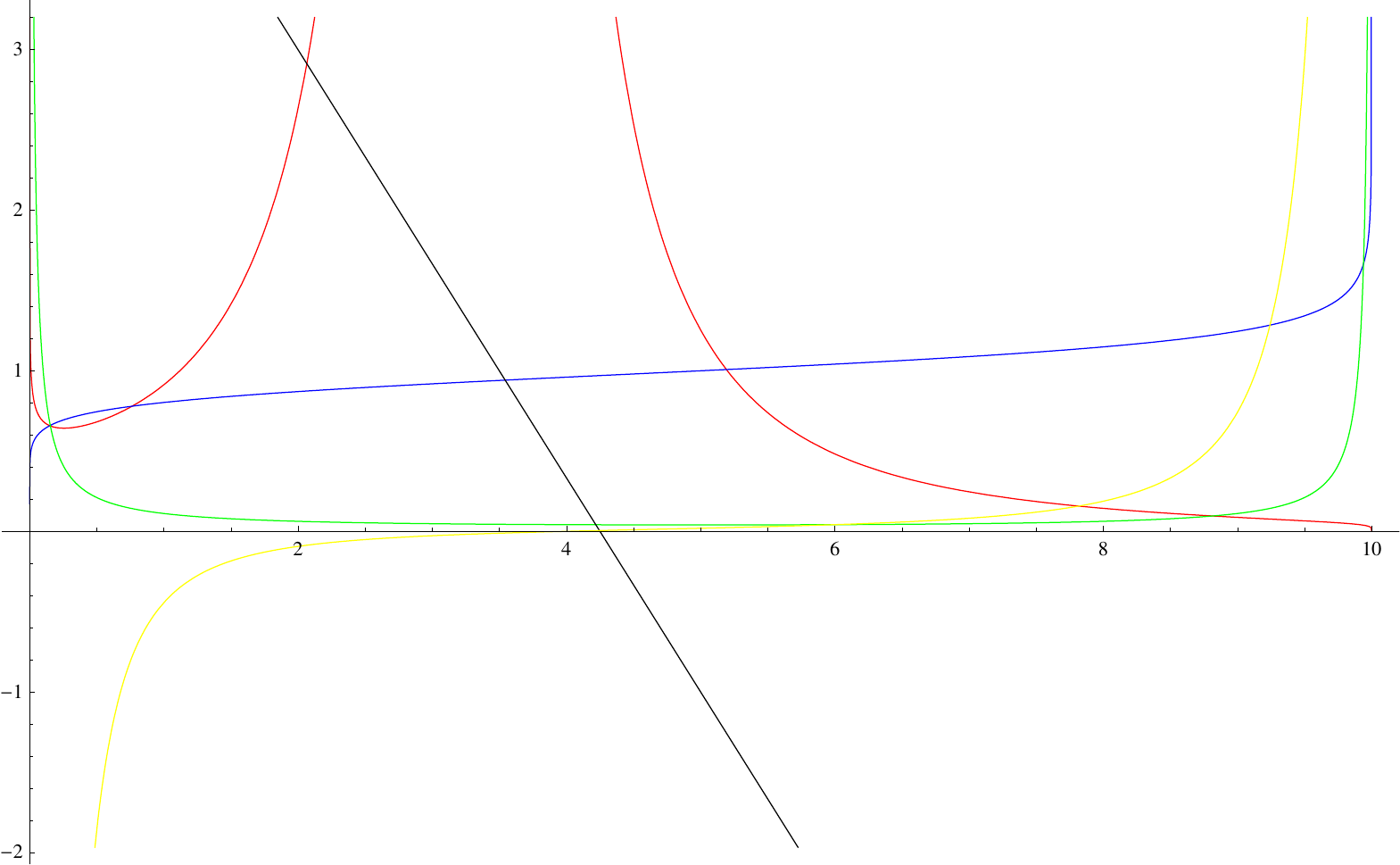}
\vspace{-5mm}
\caption{scale factor (blue line), the Hubble rate  (green line), function $u$ (red line), the effective parameter  EoS (black line) and  the scalar curvature constructed from the metric (\ref{eaa1}) (yellow line)as a function of $t$
with the parameters $h_0=0.1$, $t_0=10$, $c=2$  and $a_0=0.1$.}
\label{graph_1}
\end{figure}

\section*{Acknowledgments}

This work is  supported by
the grant of Russian Ministry of Education and Science, project TSPU-139.


\begin{thebibliography}{99}

\bibitem{review}
S. Nojiri and S.~D.~Odintsov,
   eConf C {\bf 0602061} (2006) 06
[Int.\ J.\ Geom.\ Meth.\ Mod.\ Phys.\  {\bf 4} (2007) 115] [ hep-th/0601213];
Phys.\ Rept.\  {\bf 505} (2011) 59  [arXiv:1011.0544 [gr-qc]]; Int.\ J.\ Geom.\
Meth.\ Mod.\ Phys.\  {\bf 11} (2014) 1460006
   [arXiv:1306.4426 [gr-qc]];
V.~Faraoni and S.~Capozziello,
``Beyond Einstein gravity : A Survey of gravitational theories for cosmology
and astrophysics,''
Fundamental Theories of Physics, Vol. 170, Springer, 2010;
S.~Capozziello and
M.~De Laurentis,
   Phys.\ Rept.\  {\bf 509} (2011) 167
   [arXiv:1108.6266 [gr-qc]];  G.~J.~Olmo,
  Int.\ J.\ Mod.\ Phys.\ D {\bf 20}, 413 (2011)
  [arXiv:1101.3864 [gr-qc]].


\bibitem{R}
  S.~Nojiri and S.~D.~Odintsov,
   Phys.\ Rev.\ D {\bf 68} (2003) 123512
   [hep-th/0307288].

\bibitem{BI}
S. Deser and G. W. Gibbons, Class. Quant. Grav. {\bf 15} (1998) L35;
M.
Ba\~nados and P. G. Ferreira, Phys. Rev. Lett. {\bf 105} (2010) 011101.


\bibitem{Nojiri:2007as}
 S.~'i.~Nojiri and S.~D.~Odintsov,
  Phys.\ Lett.\ B {\bf 657} (2007) 238
[arXiv:0707.1941 [hep-th]];
>    
 S.~'i.~Nojiri and S.~D.~Odintsov,
 Phys.\ Rev.\ D {\bf 77} (2008) 026007
[arXiv:0710.1738 [hep-th]];
G .~Cognola, E.~Elizalde, S.~Nojiri, S.~D.~Odintsov, L.~Sebastiani and
 S.~Zerbini,
Phys.\ Rev.\ D {\bf 77} (2008) 046009
[arXiv:0712.4017 [hep-th]];
 E.~Elizalde, S.~Nojiri, S.~D.~Odintsov, L.~Sebastiani and S.~Zerbini,
Phys.\ Rev.\ D {\bf 83} (2011) 086006
 [arXiv:1012.2280 [hep-th]].



\bibitem{Eb6}
A.N. Makarenko, S. Odintsov, G.J. Olmo,
[arXiv:1403.7409 [hep-th]].

\bibitem{Eb7}
A.N. Makarenko, S. Odintsov, G.J. Olmo,
Phys.Lett. B {\bf 734} (2014) 36
[arXiv:1404.2850 [gr-qc]].


\bibitem{OD1}
  S.~Nojiri and S.~D.~Odintsov,
Gen.Rel.Grav. {\bf 38} (2006) 1285
      [hep-th/0506212].







%





\end{thebibliography}
\end{document}